\documentclass{PoS}

\usepackage[utf8]{inputenc}
\usepackage{amsmath}
\usepackage{amsfonts}
\usepackage{amssymb}
\usepackage{graphicx}
\usepackage{braket}
\usepackage{bbm}
\usepackage{wrapfig}

\newcommand{\Obs}{{\cal O}}

\def\sl3c{\text{SL}(3,\mathbb{C})}
\def\su3{\text{SU(3)}}
\newcommand{\comment}[1]{}
\newcommand{\parset}{\xi}
\newcommand{\figscale}{0.65}

\title{Reweighted complex Langevin and its application to two-dimensional QCD}
\ShortTitle{Reweighted complex Langevin and its application to two-dimensional QCD}

\author{\speaker{Jacques Bloch}\\
        University of Regensburg\\
        E-mail: \email{jacques.bloch@ur.de}}
        
\author{Johannes Meisinger\\
        University of Regensburg\\
        E-mail: \email{johannes.meisinger@ur.de}}

\author{Sebastian Schmalzbauer\\
        Goethe University of Frankfurt \\
        E-mail: \email{schmalzbauer@th.physik.uni-frankfurt.de}}

\dedicated{Supported by the Deutsche Forschungsgemeinschaft (SFB/TRR-55)}

\FullConference{34th annual International Symposium on Lattice Field Theory\\
		24-30 July 2016\\
		University of Southampton, UK}

\abstract{
We present the reweighted complex Langevin method, which enlarges the applicability range of the complex Langevin method by reweighting the complex trajectories. In this reweighting procedure both the auxiliary and target ensembles have a complex action. We validate the method by applying it to two-dimensional strong-coupling QCD at nonzero chemical potential, and observe that it gives access to parameter regions that could otherwise not be reached with the complex Langevin method.
}

\begin{document}

\section{Introduction}

A major obstacle to simulate lattice QCD at nonzero chemical potential is formed by the complex valued determinant of the Dirac operator. This \emph{sign problem} prohibits the use of importance sampling algorithms to generate relevant configurations and most solutions to circumvent this problem have a computational cost that grows exponentially in the volume and are restricted to the region of the phase diagram where $\mu/T<1$, away from the critical region.

An alternative which became increasingly popular in recent years is the complex Langevin (CL) method. Although the equations employed in this method are straightforward generalizations of the real Langevin equations from real to complex actions, the validity of the complexified equations require a number of conditions to be met.
It is important to investigate theories and models with complex actions to deepen our understanding when the complex Langevin method can be used and trusted. Recent investigations in heavy-dense QCD \cite{Sexty:2013ica} and full QCD \cite{Fodor:2015doa} showed that the method breaks down in the transition region. Problems with the CL method were also uncovered in low-dimensional strong-coupling QCD \cite{Bloch:2015coa}, as the method converges to wrong values for small masses at large coupling.

In this presentation we introduce a novel idea where we combine the CL method and reweighting of complex trajectories to form the reweighted complex Langevin (RCL) method \cite{Bloch2016}. As we will show in the results on two-dimensional QCD this method allows us to reach regions of parameter space that are not simulated correctly by the CL method alone.

\section{Complex Langevin Method}

We first briefly introduce the complex Langevin method. Assume a partition function
\begin{align}
Z = \int dx \, e^{-S(x)} ,
\label{partfunc}
\end{align}
with real degrees of freedom $x$ and complex action $S(x)$, where $x$ can be assumed to a multidimensional vector of variables. 
When applying the Langevin equations on a system with a complex action the real variables are automatically driven into the complex plane, such that $x \to z = x + i y$. These complex variables satisfy the CL evolution equation
\begin{align}
\dot z(t) = -\frac{\partial S}{\partial z} + \eta(t) .
\end{align}
This equation can be solved numerically after proper discretization and the standard stochastic Euler discretization yields the discrete Langevin time evolution
\begin{align}
z(t+1) = z(t) + \epsilon K + \sqrt{\epsilon}\,\eta ,
\end{align}
with drift $K = -\partial S/\partial z$, step size $\epsilon$ and
independent Gaussian noise $\eta$ (chosen real for better convergence) with mean 0 and variance 2.

It is important to understand if and when the fixed point solution of the complex Langevin equation reproduces the correct expectation values of the partition function \eqref{partfunc}.
It was shown that if the action $S$ and the observable ${\cal O}$ are holomorphic in the complexified variables (up to singularities)  the crucial equivalence identity
\begin{align}
\braket{\Obs} \equiv  \frac{1}{Z}\int dx \, e^{-S(x)} {\cal O}(x) = \int dx dy \, P(x+iy) {\cal O}(x+iy) 
\label{equiv}
\end{align}
holds if the probability density $P(z)$ of the complexified variables $z$ along the CL trajectories is suppressed close to singularities of drift and observable and decays sufficiently rapidly in the imaginary direction of $z$ \cite{Aarts:2011ax,Nagata:2016vkn}.
If the CL validity conditions are not satisfied for some parameter values the CL method will fail and produce incorrect results. For QCD this depends on the values of the parameters $\mu$, $m$, $\beta$, see Sec.\ \ref{QCD}.

\section{Reweighting the complex Langevin trajectories}

Below we introduce the reweighted complex Langevin (RCL) method \cite{Bloch2016}. The principle of the method is to generate a CL trajectory for an auxiliary ensemble where the CL method is valid and to reweight this complex trajectory to compute observables in a target ensemble. The aim is to extend the applicability range of the CL method to parameter regions for which the CL validity conditions may not be satisfied.

Consider a target ensemble with parameters $\parset$ and an auxiliary ensemble with parameters $\parset_0$. The general reweighting formula to compute expectation values in the target ensemble using configurations from the auxiliary ensemble is given by
\begin{align}
\braket{\Obs}_{\parset} &= \frac{\int dx \, w(x;\parset) \Obs(x;\parset)}{\int dx \,w(x;\parset)} 
= \frac{\int dx \, w(x;\parset_0) \left[\frac{w(x;\parset)}{w(x;\parset_0)}\Obs(x;\parset)\right]}{\int dx \,w(x;\parset_0) \left[\frac{w(x;\parset)}{w(x;\parset_0)}\right]} 
= \frac{\Braket{\frac{w(x;\parset)}{w(x;\parset_0)}\Obs(x;{\parset})}_{\parset_0}}{\Braket{\frac{w(x;\parset)}{w(x;\parset_0)}}_{\parset_0}} .
\label{reweighting}
\end{align}
The peculiarity of our reweighting method is that we consider an auxiliary ensemble at nonzero chemical potential such that the auxiliary weights $w(x;\parset_0)$ are complex, whereas these are taken real and positive in standard reweighting procedures. 
Therefore we cannot use importance sampling to generate relevant configurations in the auxiliary ensemble, but will instead use the CL method to generate an auxiliary trajectory of complex valued configurations.

If the CL method is valid for the parameters $\parset_0$, the CL equivalence \eqref{equiv} holds for the auxiliary ensemble and can be applied to both $\Braket{\cdots}_{\parset_0}$ in the reweighting formula \eqref{reweighting}, which yields the following RCL equation:
\begin{align}
\braket{\Obs}_{\parset} 
&= \frac{\int dx dy \,  P(z;\parset_0) \left[\frac{w(z;\parset)}{w(z;\parset_0)}\Obs(z;\parset)\right]}{\int dx dy \, P(z;\parset_0) \left[\frac{w(z;\parset)}{w(z;\parset_0)}\right]} .
\label{RCL}
\end{align}
The expectation value $\braket{\Obs}_{\parset}$ in the target ensemble is thus computed as a ratio of averages, both evaluated along the auxiliary CL trajectory. 
As the CL method is valid in the auxiliary ensemble,  both expectation values in this ratio will be  evaluated reliably, independently of the fact if the CL method itself is valid or not in the target ensemble. Indeed, as we will see in the next section, the RCL method also works well if the CL validity conditions are violated in the target ensemble.

As this reweighting along complex trajectories is a novel idea, it is important to put the method at a test and verify that it works as expected. In the next section we show results in two-dimensional QCD, but the method has also been successfully tested on random matrix models for QCD.

\section{Reweighted complex Langevin for 1+1d QCD}
\label{QCD}

\subsection{Complex Langevin for QCD}

Let us consider the partition function of lattice QCD,
\begin{align}
Z=\left[\prod_{x=1}^V\prod_{\nu=0}^{d-1} \int\! d U_{x,\nu}\right]\, \exp[-S_g(\beta)] \det D(m;\mu) ,
\end{align}
with SU(3) matrices $U_{x,\nu}$, Wilson gauge action $S_g(\beta)$ and staggered Dirac operator $D(m;\mu)$ for a quark of mass $m$ at chemical potential $\mu$. For nonzero $\mu$ the Dirac determinant is complex, so we have a complex action and a potential sign problem. In this case, the CL equations will drive the links from SU(3) into $\sl3c$. When the complex trajectories wander off too far from SU(3) the CL method becomes invalid. This problem is resolved in gauge theories by applying \emph{gauge cooling} \cite{Seiler:2012wz}, which keeps the trajectories as close as possible to SU(3) using $\sl3c$ gauge invariance. 

Although gauge cooling can alter the CL trajectory such that the CL validity conditions are satisfied, there is no guarantee that this can be  achieved for all parameter values. This was investigated in detail for strong-coupling QCD in 1+1 dimensions \cite{Bloch:2015coa} where we showed that, even with gauge cooling, the CL results are only valid in a certain ($m,\mu$) range. At small masses the CL trajectories still populate the singularity at $\det D=0$ and CL irrecoverably gives wrong results.
These are cases of interest to verify if the RCL method can be used to recover the correct results.

\subsection{Reweighting in the mass}

\begin{figure}[b]
\centerline{\includegraphics[scale=\figscale]{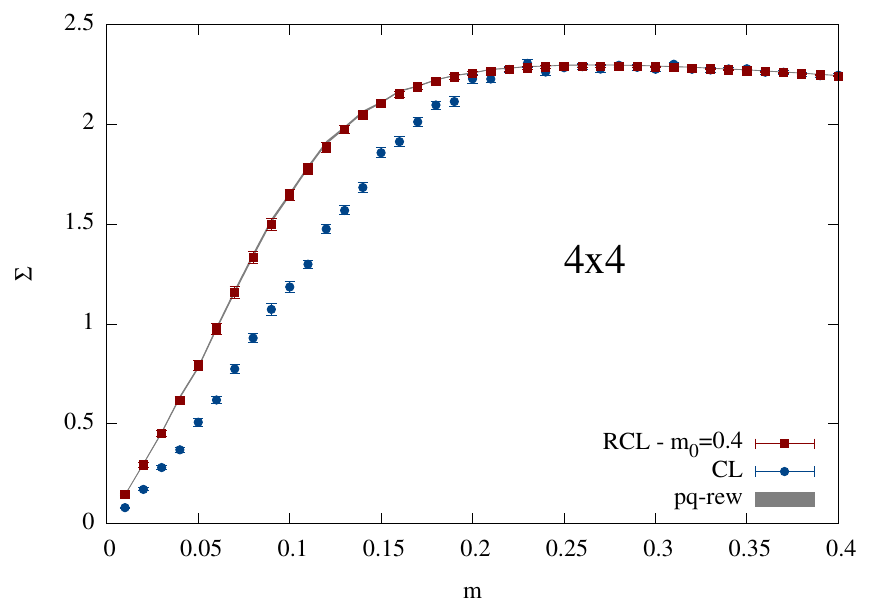}
\includegraphics[scale=\figscale]{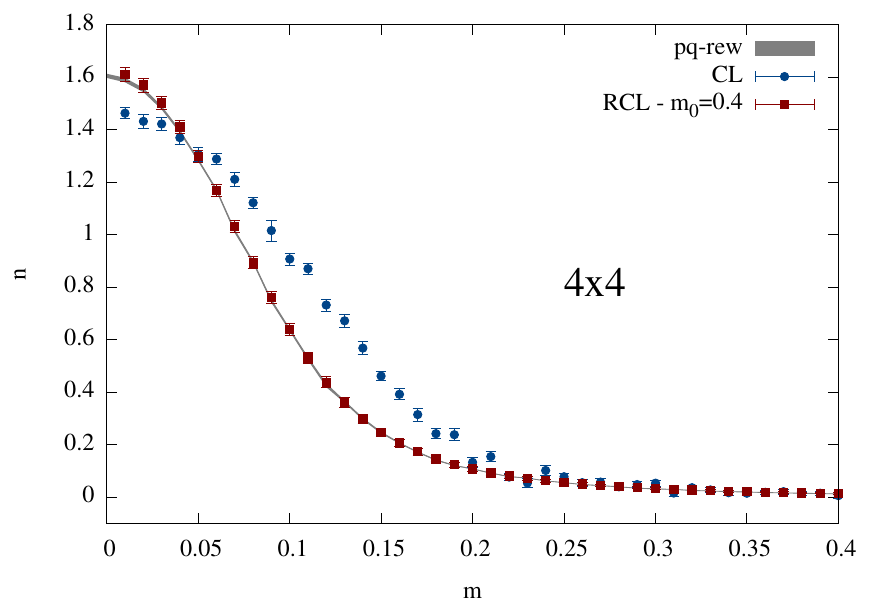}}
\centerline{\includegraphics[scale=\figscale]{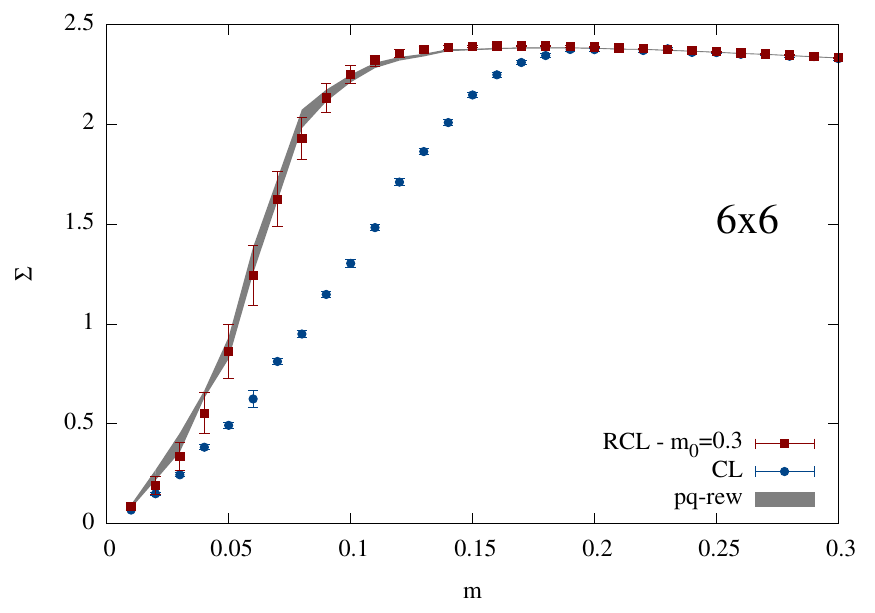}
\includegraphics[scale=\figscale]{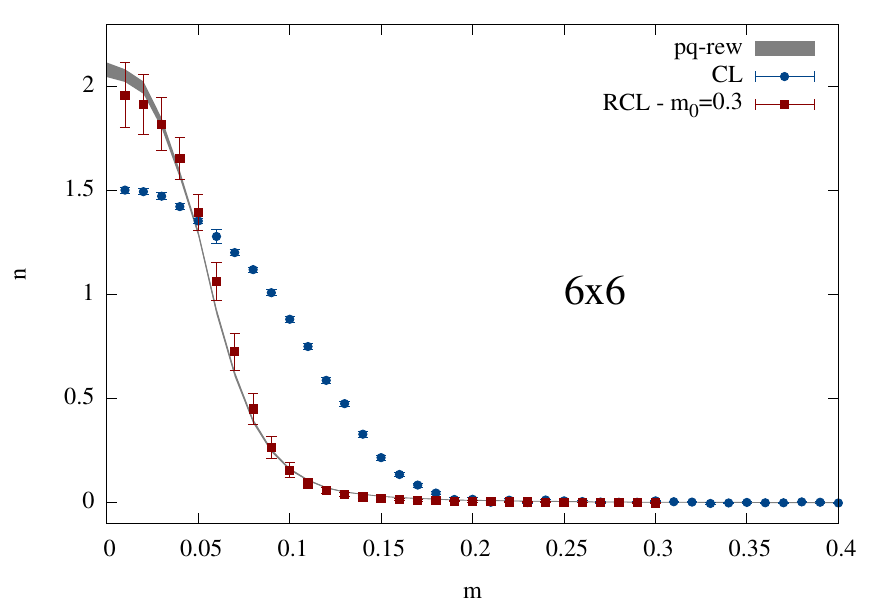}}
\caption{Chiral condensate (left) and number density (right) versus mass at $\mu=0.3$ for a $4\times 4$ lattice (top row) and a $6\times 6$ lattice (bottom row): data for CL (blue points) versus RCL (red points). The grey band are benchmark results computed with standard phase-quenched reweighting.} 
\label{RCL-m}
\end{figure}

We first test the CL method by computing the chiral condensate and the number density on a $4\times4$ lattice in the strong-coupling limit $\beta=0$ and choose $\mu=0.3$ as the sign problem is most pronounced around this value. Even though the sign problem is still mild on such a small lattice, the results of gauge cooled CL are wrong for small masses ($m\lesssim0.2$), as is illustrated by the blue points in the top row of Fig.\ \ref{RCL-m}, due to the singular drift problem. The benchmark results (grey band) were computed with phase-quenched reweighting simulations. We apply the RCL equation \eqref{RCL} using an auxiliary CL trajectory generated at mass $m_0=0.4$ for the same $\mu$ and compute the RCL results for the range $m\in[0,0.4]$. The results are given by the red points in the figure. Clearly RCL in mass works over the complete mass range on this lattice size, even in the strong coupling limit.

Next we tested the RCL method on a $6\times6$ lattice at $\beta=0$, again for $\mu=0.3$ where the sign problem is strongest. The results are shown in the bottom row of Fig.\ \ref{RCL-m}. Again, gauge cooled CL (blue points) is wrong for small masses ($m\lesssim0.2$). The RCL method uses an auxiliary trajectory generated at $m_0=0.4$ and $\mu_0=\mu$ to compute the results for $m\in[0,0.4]$. On this somewhat larger lattice the overlap and sign problems become visible through larger error bars, but nevertheless, the RCL method does work down to very small masses.

\subsection{Reweighting in the chemical potential}

\begin{figure}[b]
\centerline{\includegraphics[scale=\figscale]{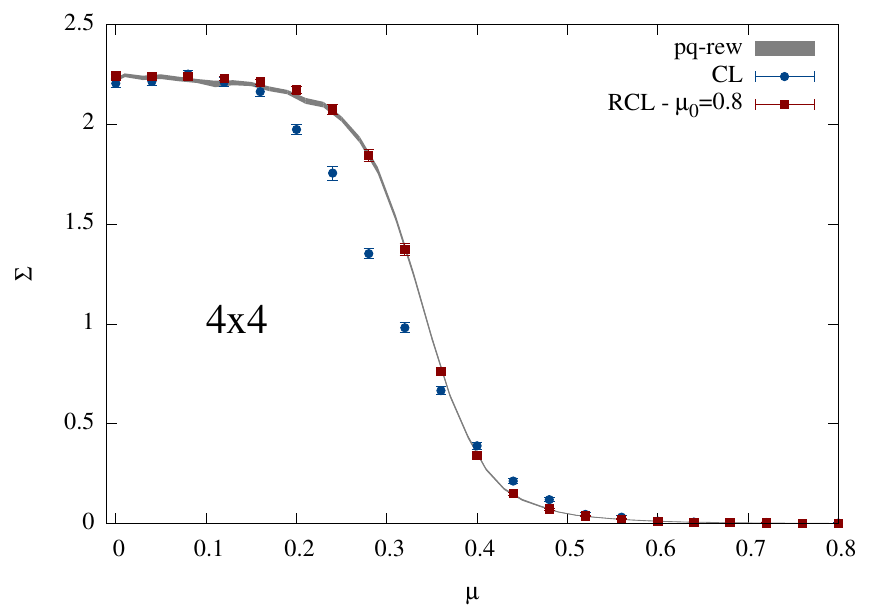}
\includegraphics[scale=\figscale]{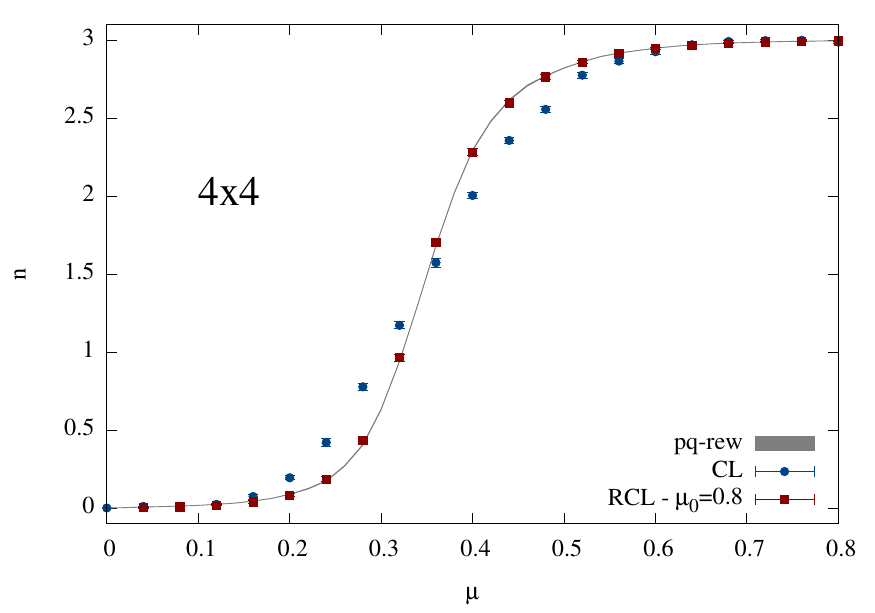}}
\centerline{\includegraphics[scale=\figscale]{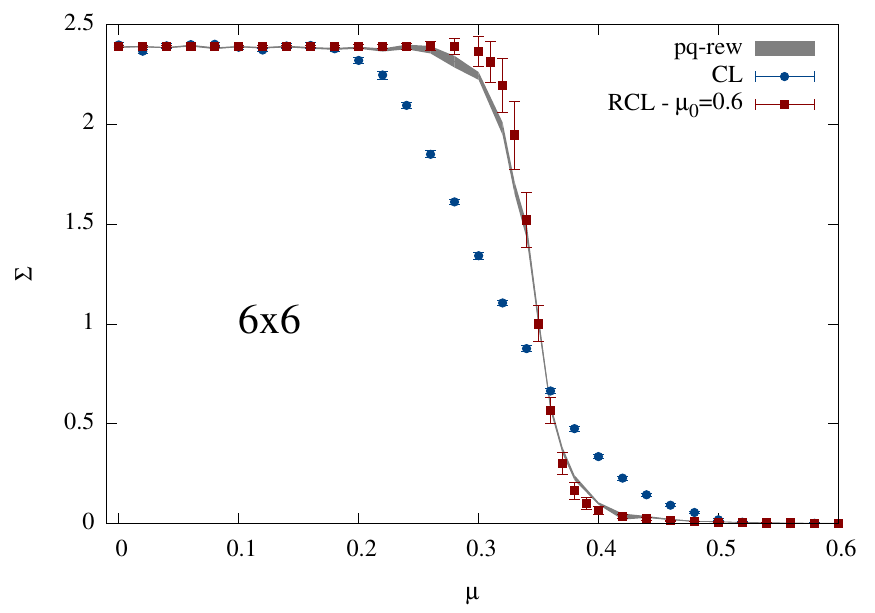}
\includegraphics[scale=\figscale]{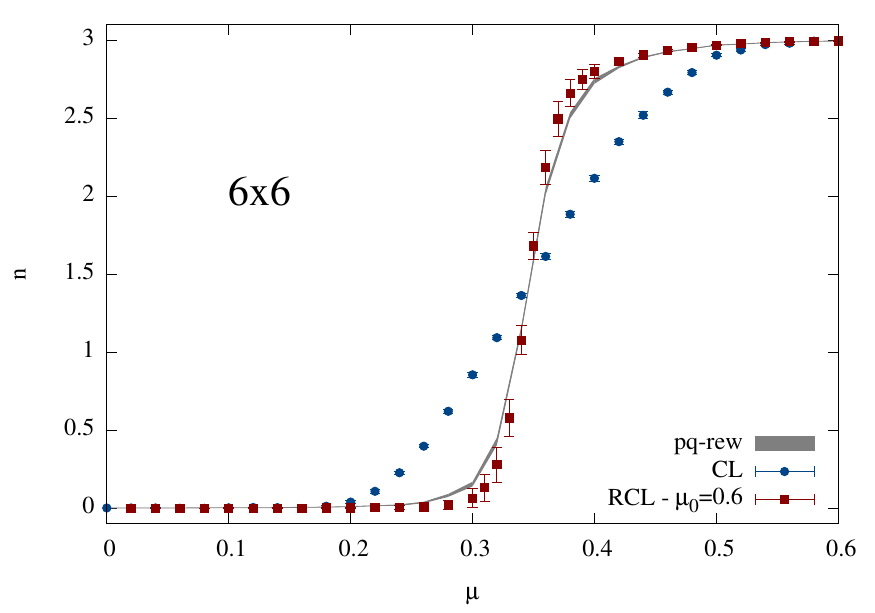}}
\caption{Chiral condensate (left) and quark number (right) for $m=0.1$ on $4\times4$ (top) and $6\times6$ (bottom) lattices versus chemical potential $\mu$. Results for CL (blue points) and RCL (red points).  The grey band benchmark is computed using phase-quenched reweighting.}
\label{RCL-mu}
\end{figure}

Clearly we can use the RCL method to reweight in any relevant parameter and we here investigate the reweighting in the chemical potential $\mu$. In Fig.\ \ref{RCL-mu} we show the strong-coupling ($\beta=0$) results for small mass $m=0.1$ on $4\times4$ (top) and $6\times6$ (bottom) lattices as a function of $\mu$. We see that for such small masses the gauge cooled CL method fails to reproduce the correct results. The RCL results are computed using an auxiliary CL trajectory generated at $\mu_0=0.8$ for the $4\times4$ lattice and at $\mu_0=0.6$ for the $6\times6$ lattice. In all cases the RCL results agree with the benchmark over the complete $\mu$ range within the error bars. The increasing errors for the $6\times6$ lattice in the critical region point to an increasing sign problem. It is however surprising that RCL works well for small $\mu$, i.e., far from the auxiliary value, meaning that there is no serious overlap problem in this case. When performing this reweighting in $\mu$ for even smaller masses the reweighting method will start to break down as the sign problem becomes stronger.

\subsection{Reweighting in the coupling}

Even though RCL in $\mu$ and $m$ could just as well have been performed at nonzero $\beta$, away from the strong coupling limit, we chose to work at $\beta=0$ as this pushes the CL method to its limits.

As a last test we use RCL to perform reweighting in $\beta$. It is known that the CL method is valid for large $\beta$ but breaks down for lower values. Simulations at such $\beta$ values are however needed to reach the critical region in current lattice simulations. It would therefore be helpful if RCL could be applied to valid CL trajectories generated at large $\beta$ to reach lower $\beta$ values. 

\begin{figure}[h]
\centerline{\includegraphics[scale=\figscale]{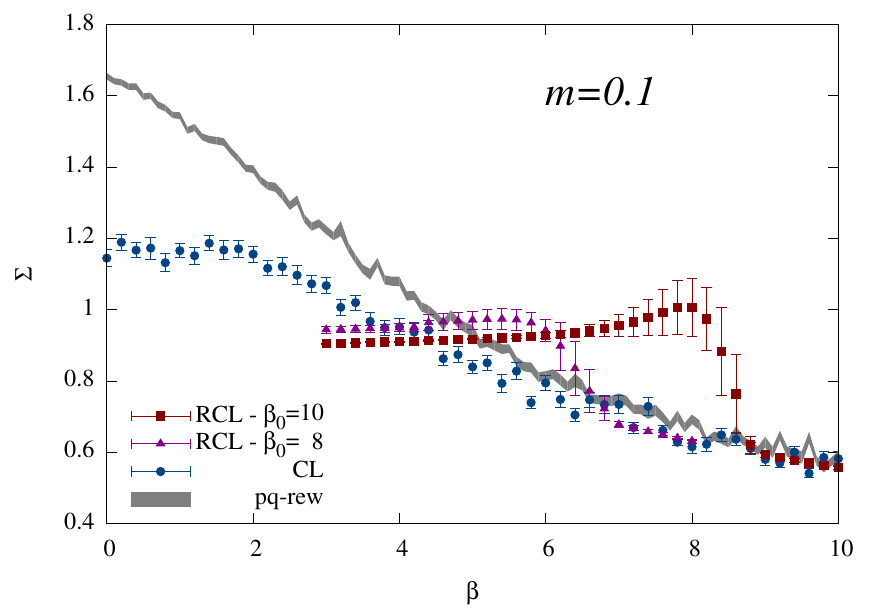}\vspace{-2mm}}
\caption{Chiral condensate versus $\beta$ on a $4\times 4$ lattice for $\mu=0.3$ and $m=0.1$. Results for CL (blue points) versus RCL (red and purple points).}
\label{RCL-beta}
\end{figure}

Fig.\ \ref{RCL-beta} shows the results for $\mu=0.3$ and $m=0.1$ on a $4\times4$ lattice. The CL method only works correctly for large $\beta>6$, so we investigate if RCL allows us to reach lower $\beta$ values. Unfortunately, the range of application of RCL in $\beta$ seems quite restricted: starting from an auxiliary trajectory at $\beta_0=10$ RCL works down to $\beta\approx8.5$ and from $\beta_0=8$ it works down to $\beta\approx6.5$. In both cases RCL does not perform better than the original CL. This is due to the extremely sharp probability density of the gauge action, where $\beta$ is a multiplicative factor in the exponential. Even tiny changes in $\beta$ strongly affect the gauge weight and reweighting is inefficient.

\section{Some additional remarks}

Although RCL works correctly to reweight from one set of parameters to another, it suffers from the usual overlap and sign problems. 
A possible advantage over other reweighting procedures (phase-quenched, Glasgow and quenched reweightings) is that the auxiliary ensemble at $\mu\neq 0$ could be closer to the target ensemble, thus increasing the overlap between the target and auxiliary ensembles.
In Glasgow reweighting the auxiliary ensemble is taken at $\mu_0=0$. Clearly, reweighting from $\mu_0 \neq 0$ using RCL starts from an auxiliary ensemble that is closer to the target ensemble.

In phase-quenched reweighting the auxiliary ensemble uses the magnitude of the fermion determinant as sampling weights. The auxiliary and target ensembles are in different phases when $\mu>m_\pi/2$ and there is therefore little overlap between the relevant configurations in both ensembles. In RCL, however, the auxiliary and target ensembles are both taken in full QCD and hence this problem could be alleviated. Moreover,  RCL uses a single CL trajectory to reweight to a range of target parameter values, whereas phase-quenched reweighting typically constructs a new Markov chain for each new parameter value.

\section{Summary and outlook}

For many theories with a complex action the complex Langevin method works correctly for some range of parameters, but fails for other parameter values when the validity conditions are violated. In this talk we have presented the \textit{reweighted complex Langevin} method, which combines complex Langevin and reweighting to compute observables in a target ensemble using complex trajectories generated for an auxiliary ensemble for which the CL validity conditions are met.

As a proof of principle we applied RCL on QCD in 1+1 dimension using reweighting in $m$, $\mu$ and $\beta$ at $\mu\neq 0$ and verified that the RCL procedure works correctly. We observed that RCL works best when reweighting in the mass, while reweighting in $\mu$ works well as long as the mass is not too small. Reweighting in $\beta$ hardly works as the gauge probability is narrow and very sensitive to $\beta$. Clearly, the method could be further optimized by making a multiparameter RCL in $\mu$, $m$, $\beta$ \cite{Fodor:2001au}.
As the method suffers from the usual overlap and sign problems, its efficacy should be investigated further.

As an outlook for future work we can pinpoint a couple of avenues. One interesting application would be to test RCL on full four-dimensional QCD where it was shown that CL breaks down in the critical region. As mass reweighting works best the strategy could be to choose a high enough $m$ to get a valid CL trajectory for a particular $(\mu,\beta)$ and then reweight in $m$ to get down to the physical mass region. Alternatively one could follow a line in the $(m,\mu)$-plane keeping $\beta$ fixed. 

Clearly, one still has to learn how to reweight most efficiently with the RCL method. A useful exercise would be to make a validity map of the CL method in the $(m,\mu,\beta)$-space for two-dimensional QCD and devise the best reweighting path to cover all parameter values.

Note that RCL opens a new avenue as reweighting in the chemical potential can be extended to interpolate rather then just extrapolate if we use auxiliary ensembles at $\mu_0$ values above and below the critical region, which could improve the quality of the results.

\bibliographystyle{jbJHEP_notitle.bst}
\bibliography{biblio.bib}

\providecommand{\href}[2]{#2}\begingroup\raggedright\begin{thebibliography}{1}

\bibitem{Sexty:2013ica}
D.~Sexty,\href{http://dx.doi.org/10.1016/j.physletb.2014.01.019}{ {\em Phys.
  Lett.} {\bf B729} (2014) 108} [\href{http://arxiv.org/abs/1307.7748}{{\tt
  arXiv:1307.7748}}].

\bibitem{Fodor:2015doa}
Z.~Fodor, S.~D. Katz, D.~Sexty, and
  C.~Török,\href{http://dx.doi.org/10.1103/PhysRevD.92.094516}{ {\em Phys.
  Rev.} {\bf D92} (2015) 094516} [\href{http://arxiv.org/abs/1508.05260}{{\tt
  arXiv:1508.05260}}].

\bibitem{Bloch:2015coa}
J.~Bloch, J.~Mahr, and S.~Schmalzbauer, {\em PoS} {\bf LATTICE2015} (2016) 158
  [\href{http://arxiv.org/abs/1508.05252}{{\tt arXiv:1508.05252}}].

\bibitem{Bloch2016}
J.~Bloch, \href{http://arxiv.org/abs/1701.00986}{{\tt arXiv:1701.00986}}.

\bibitem{Aarts:2011ax}
G.~Aarts, F.~A. James, E.~Seiler, and I.-O.
  Stamatescu,\href{http://dx.doi.org/10.1140/epjc/s10052-011-1756-5}{ {\em Eur.
  Phys. J.} {\bf C71} (2011) 1756} [\href{http://arxiv.org/abs/1101.3270}{{\tt
  arXiv:1101.3270}}].

\bibitem{Nagata:2016vkn}
K.~Nagata, J.~Nishimura, and
  S.~Shimasaki,\href{http://dx.doi.org/10.1103/PhysRevD.94.114515}{ {\em Phys.
  Rev.} {\bf D94} (2016) 114515} [\href{http://arxiv.org/abs/1606.07627}{{\tt
  arXiv:1606.07627}}].

\bibitem{Seiler:2012wz}
E.~Seiler, D.~Sexty, and I.-O.
  Stamatescu,\href{http://dx.doi.org/10.1016/j.physletb.2013.04.062}{ {\em
  Phys. Lett.} {\bf B723} (2013) 213}
  [\href{http://arxiv.org/abs/1211.3709}{{\tt arXiv:1211.3709}}].

\bibitem{Fodor:2001au}
Z.~Fodor and S.~Katz,\href{http://dx.doi.org/10.1016/S0370-2693(02)01583-6}{
  {\em Phys. Lett.} {\bf B534} (2002) 87}
  [\href{http://arxiv.org/abs/hep-lat/0104001}{{\tt hep-lat/0104001}}].

\end{thebibliography}\endgroup

\end{document}